\documentclass[structabstract]{aa}
\usepackage{txfonts}
\usepackage{natbib}
\usepackage{epsfig}
\usepackage{graphicx}
\usepackage{}
\usepackage{subfigure}
\usepackage{latexsym}
\usepackage{pifont}
\input epsf

\newcommand{\xmm}{{\it XMM~\/}}
\newcommand{\xmmn}{{\it XMM-Newton~\/}}

\newcommand{\chandra}{{\it Chandra~\/}}

\newcommand{\pwn}{{pulsar wind nebula~\/}}
\newcommand{\obs}{{observation~\/}}
\newcommand{\obss}{{observations~\/}}

\def\ergcms{{\rm ~erg~cm^{-2}~s^{-1}}}

\def\ergsec{{\rm ~erg~s^{-1}}}
\def\cm2{{\rm ~cm^{-2}}}

\def\H0{{\rm ~km~s^{-1}~Mpc^{-1}}}

\def\ie{{\it i.e.~\/}}

\def\la{\mathrel{\hbox{\rlap{\hbox{\lower4pt\hbox{$\sim$}}}{\raise2pt\hbox{$<$}}}}}
\def\ga{\mathrel{\hbox{\rlap{\hbox{\lower4pt\hbox{$\sim$}}}{\raise2pt\hbox{$>$}}}}}
\def\d25{D$_{25}$}

\def\lx{L$_{\rm X}$}
\def\fx{F$_{\rm X}$}
\def\deg{\hbox{$^\circ$}}
\def\arcm{\hbox{$^\prime$~\/}}
\def\arcs{\hbox{$^{\prime\prime}$}}

\def\eps@scaling{1.0}%
\newcommand\plottwo[2]{%
  \centering
  \leavevmode
  \columnwidth=.45\columnwidth
  \includegraphics[width={\eps@scaling\columnwidth}]{#1}%
  \hfil
  \includegraphics[width={\eps@scaling\columnwidth}]{#2}%
}%

\begin{document}

\title{IKT 16: 
the first X-ray confirmed composite SNR in the SMC}

\author{C.~Maitra\inst{1}
\and J.~Ballet\inst{1}
\and M.~D.~Filipovi\'c\inst{2}
\and F.~Haberl\inst{3}
\and A.~Tiengo\inst{4,5,6}
\and K.~Grieve\inst{2}
\and Q.~Roper\inst{2}}

\institute{Laboratoire AIM, IRFU/Service d'Astrophysique - CEA/DSM - CNRS - Universite Paris Diderot, Bat. 709, CEA-Saclay, 91191 Gif-sur-Yvette Cedex, France 
\email{chandreyee.maitra@cea.fr} 
\and 
University of Western Sydney, Locked Bag 1797, Penrith South DC, NSW 1797, Australia
\and
Max-Planck-Institut f\"{u}r extraterrestriche Physik, Giessenbachstrasse, 85741 Garching, Germany
\and
IUSS, Istituto Universitario di Studi Superiori, piazza della Vittoria 15, I-27100 Pavia, Italy
\and
INAF, Istituto Nazionale di Astrofisica, IASF-Milano, via E. Bassini 15, I-20133 Milano, Italy
\and
INFN, Istituto Nazionale di Fisica Nucleare, Sezione di Pavia, via A. Bassi 6, I-27100 Pavia, Italy
} 

\date{Received ... / Accepted ...}

\abstract 
{} 
{IKT~16 is an X-ray and radio-faint supernova remnant (SNR) in the Small Magellanic Cloud (SMC). A detailed X-ray study of this SNR with \xmmn  confirmed the presence of a hard X-ray source near its centre, indicating the detection of the first composite SNR in the SMC. With a dedicated \chandra \obs we aim to resolve the point source and confirm its nature. We also acquire new ATCA observations of the source at 2.1\,GHz with improved flux density estimates and resolution.}
{We perform detailed spatial and spectral analysis of the source. With the highest resolution X-ray and radio image of the centre of the SNR available today, we resolve the source and confirm its \pwn (PWN) nature. Further, we constrain the geometrical parameters of the PWN and perform spectral analysis for the point source and the PWN separately. We also test for the radial variations of the PWN spectrum and its possible east west asymmetry.}
{The X-ray source at the centre of IKT 16 can be resolved into a symmetrical elongated feature centring a point source, the putative pulsar. Spatial modelling indicates
 an extent of 5.2\arcs of the feature with its axis inclined at 82\deg~ east from north, aligned with a larger radio feature consisting of two lobes almost symmetrical about the X-ray source. The picture is consistent with a PWN which has not yet collided with the reverse shock. The point source is about three times
brighter than the PWN and has a hard spectrum of spectral index 1.1 compared to a value 2.2 for the PWN. This points to the presence of a pulsar dominated by non-thermal emission. The expected $\dot{E}$ is $\sim 10^{37} \ergsec$ and spin period < 100 ms. However, the presence of a compact nebula unresolved by \chandra at the distance of the SMC cannot completely be ruled out. }
{}

\keywords{galaxies: Magellanic Clouds - ISM: supernova remnants}

\maketitle

\section{Introduction}
\label{sec:intro}
Composite supernova remnants (SNRs) are a subclass of core collapse SNRs where non-thermal radio emission is observed from both the expanding shell of the SNR, and from the PWN located inside it. The PWNe are powered by the relativistic outflows from the young neutron star and can be observed at all wavelengths, but mostly in the radio and X-rays.
The PWN morphology can provide crucial information on the properties of the outflow, the interacting ambient medium and the geometry of the pulsar powering it \citep[see][and references therein for detailed understanding of the structure and evolution of PWNe]{gaensler2006}. The \chandra observatory with its excellent spatial resolution and high sensitivity has provided significant breakthroughs in the study of PWNe. Apart from steadily increasing the census on the number of PWNe discovered, it has provided a unique opportunity to study in detail their spatial and spectral structures and signatures of interaction with the surrounding medium \citep{pwnchandra}. Most of these are however Galactic sources and there are no confirmed PWNe in our satellite galaxy the SMC, yet. To this day a total of 24 classified SNRs are known in the SMC \citep[][and references therein]{haberl2012}. Out of them very few are PWN candidates. One such source, HFPK 334 has been recently dismissed as a background source \citep{crawford2014}. Discovery of a PWN in the SMC will open a new window in the study of rotation powered pulsars in the satellite galaxy which has a rich history of active star formation and is therefore expected to host young energetic pulsars.

IKT 16 is an X-ray and radio-faint SNR in the SMC first studied with \xmmn in a survey of known SNRs in the SMC \citep{van2004}. In this study a region of harder 
X-ray emission was found at the centre of the remnant although it could not be probed further  due to poor statistics with a single exposure. \cite{owen2011} further used eight more archival \xmmn \obss 
(total useful exposure 125 ks and off-axis angle 8-12$\arcmin$) taken subsequently to carry out a multiwavelength study of the spatial and spectral properties of this SNR and the associated central source. The authors found substantial evidence that the unresolved source detected at the centre of the SNR by XMM-Newton, is a PWN associated with it.
Radio images from the ATCA and MOST surveys displayed faint radio structures correlated with the X-ray throughout the remnant. The brightest feature was an extended radio emission 
corresponding to the X-ray source and extending a distance of 40\arcs~ towards the centre of the SNR \citep{owen2011}. This picture is consistent with a moving pulsar in the SNR.
 Further, the large size of the remnant suggested that the SNR is in the adiabatic Sedov phase of evolution. In the Sedov model the age of the SNR was $\sim$ 14.7 kyr
and implied the PWN may have interacted with and been compressed by the reverse shock. The bright central source was located 30\arcs~ from the SNR centre implying a transverse kick 
velocity of $\sim$ 580 km s$^{-1}$. 

To further resolve the nature of the central source and establish the presence of the first PWN in the SMC, a 40 ks on-axis \chandra observation was solicited. In this paper we describe the results of the \chandra \obs of IKT 16. We have performed detailed spatial and spectral analysis of the hard X-ray emission near the centre of the SNR, previously unresolved. With the unprecedented spatial resolution of \chandra we have resolved the source into a central point source (a putative pulsar) and an extended emission surrounding it, indicating a PWN nature for the source. The \obss and analysis are described in section 2. Section 3 presents the spatial analysis including the imaging and morphological fitting of the source. Section  4 presents the detailed spectral analysis of the central source, the surrounding nebula and its decomposed regions. Section 5 presents the discussion  and section 6 the conclusions.

\begin{figure*}
\hspace*{-0.52cm}
\subfigure[]{\includegraphics[angle=0,scale=0.5]{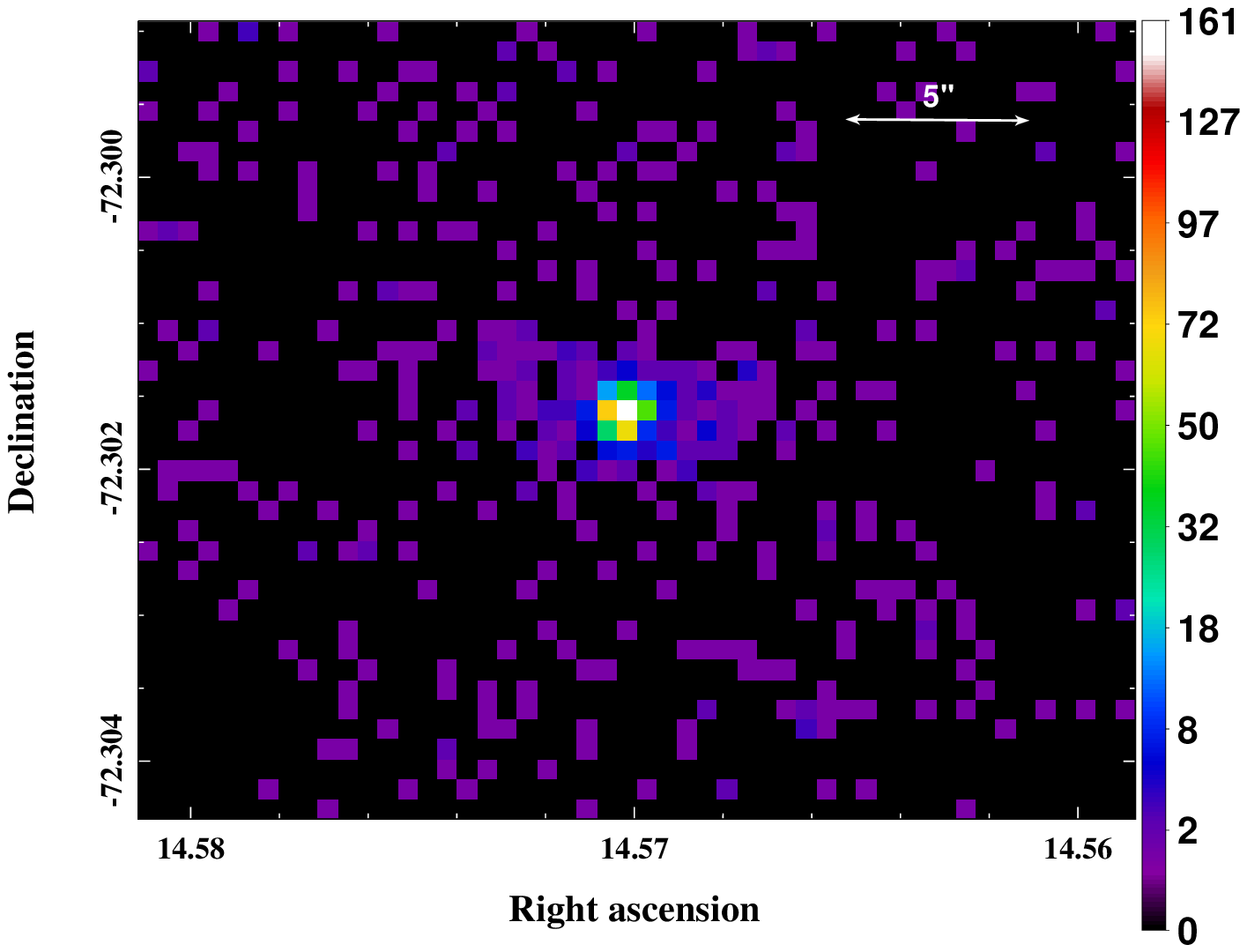}}
\hspace*{-1.5cm}
\subfigure[]{\includegraphics[angle=0,scale=0.45]{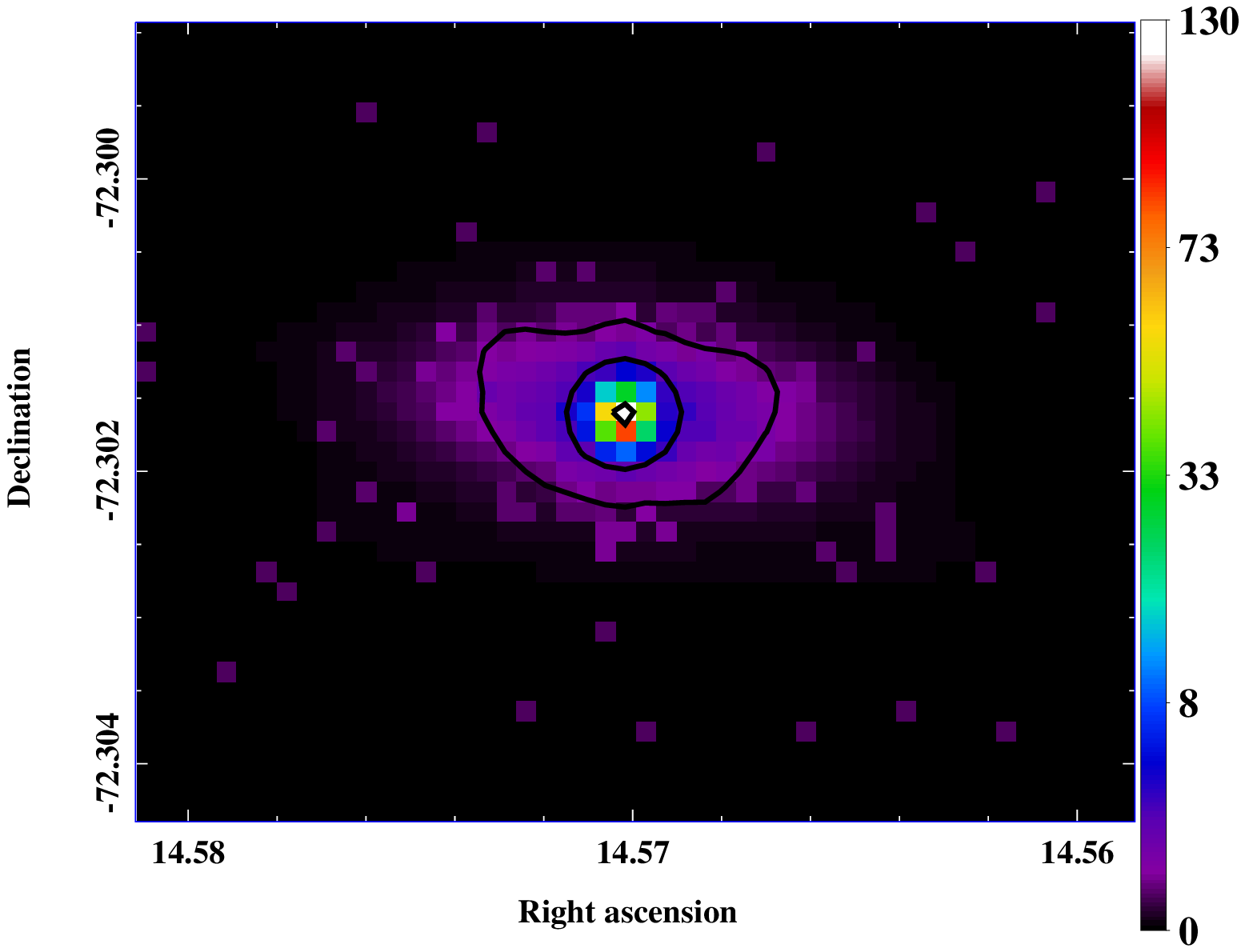}}
\subfigure[]{\includegraphics[angle=0,scale=0.52]{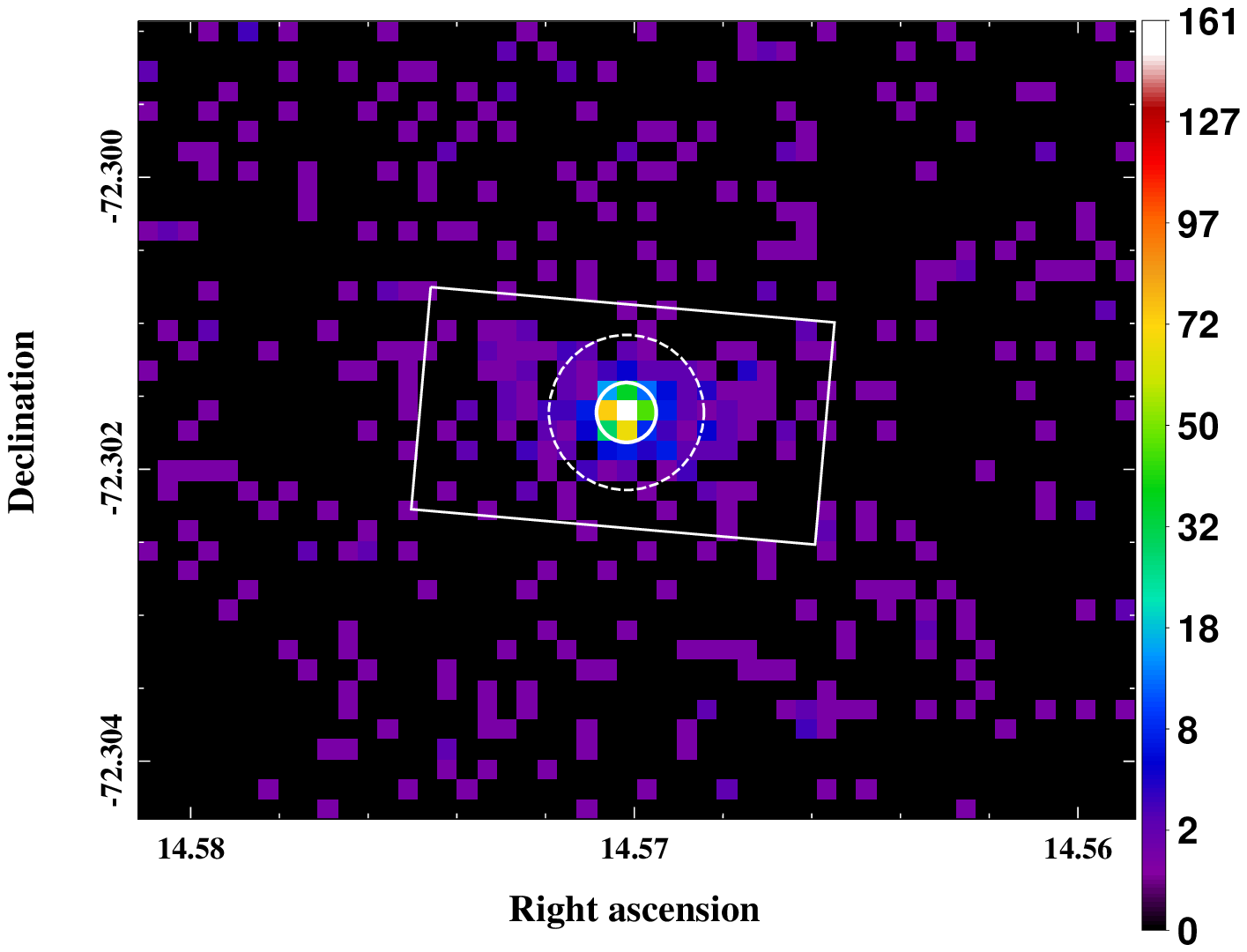}}
\vspace*{2.0cm}
\hspace*{0.18 cm}
\subfigure[]{\includegraphics[angle=0,scale=0.42]{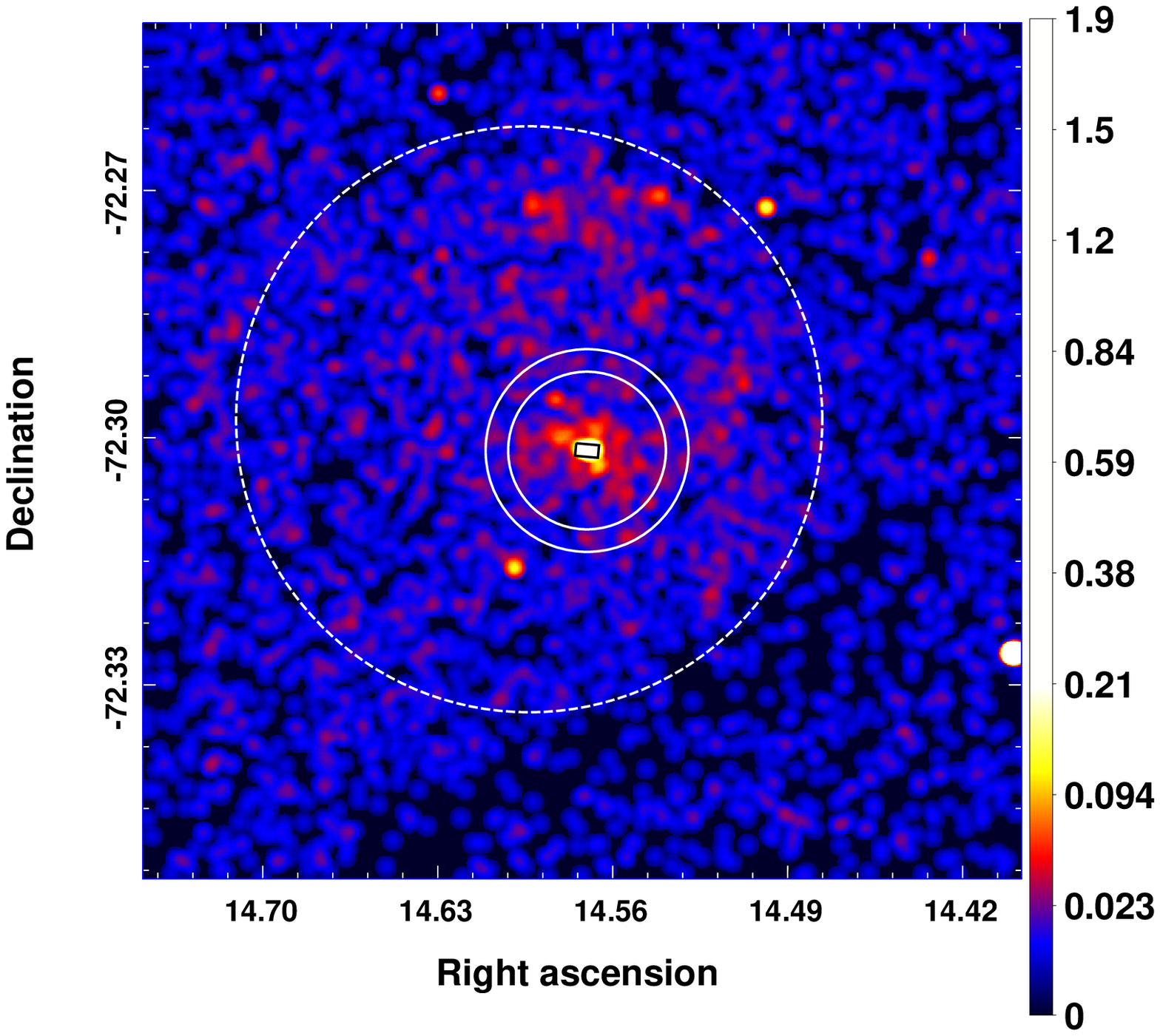}}
\caption{a({\it top left}): \chandra ACIS-S (0.5--8 keV) full resolution image of the PWN in IKT 16. The image size is 20\arcs x 25\arcs. The scale is in square root, the X and Y axis in degrees, and units are counts for all the images. The brightest pixel in the centre corresponds to the putative pulsar. b({\it top right}): Best-fit model of the PWN (0.5--8 keV). Overlayed are contours from the data which have been smoothed with a Gaussian of $\sigma = 3$\arcs. The contour levels are plotted at values of 0.7, 1.6 and 30 counts arcsecs$^{-2}$. 
c({\it bottom left}): Same as the {\it top left} figure showing the regions used for spectral extraction. The central circle corresponds to the pulsar, and the rectangular box the entire nebula with the central point source removed. The dashed circle corresponds to the outer boundary of the inner, and the inner boundary of the outer nebular extraction regions respectively. d({\it bottom right}): Larger (0.5--2 keV) image centred on the PWN of IKT 16 (ACIS-S3), with the bottom right corner of the image corresponding to ACIS-S2. The box region used for nebular extraction is shown in black solid lines and the background annular region used for spectral extraction (inner radius 34\arcs) is shown in white solid lines. The white dashed circle indicates the position and extent of the SNR in \cite{owen2011}. The point source contribution at the centre has been removed and replaced with values obtained by interpolation from the nebular region. The image has been smoothed using a Gaussian kernel of width 5\arcs.}
\label{images}
\end{figure*}

\begin{figure*}
\centering
\includegraphics[scale=0.62,angle=-90]{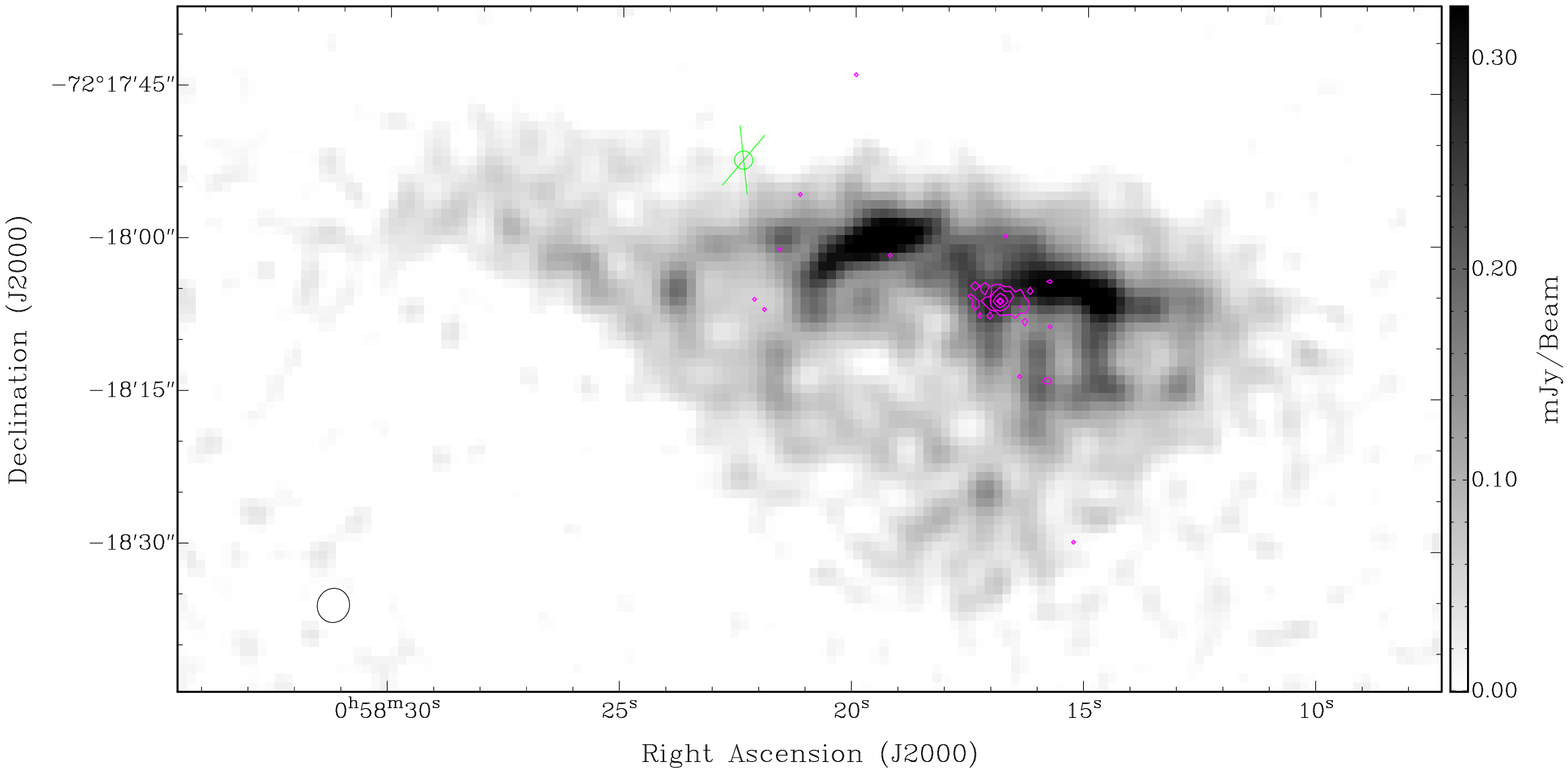}
\caption{ATCA 2.1\,GHz image of PWN IKT\,16 overlaid with Chandra contours (\textbf{1.25, 5, 25 and 100 c/arcsec$^{2}$}). The cross marked in green denotes the centre of the SNR in IKT 16.
The synthesis beam size of $3.38\arcsec \times 3.16\arcsec$ at PA of $-17.8^{\circ}$ is shown on the bottom left corner. The 2.1\,GHz image grayscale bar is shown in units of mJy/beam.}
\label{radio}
\end{figure*}


\section{Observations \& Analysis}
\label{sec:obs}

\subsection{\chandra ACIS \obs and data reduction}
\label{sec:xobs}

The \chandra \obs (ObsID 13773) was carried out with the ACIS-S as the primary instrument in the timed exposure mode. The \obs was performed on 09-02-2013 with the central source of  IKT 16 positioned at the aimpoint of the S3 CCD with an exposure of 38.5 ks. 
It is worthwhile to mention here that there was a previous \chandra \obs (obsID 2948, 9 ks) where the source was detected \citep{evans2010}. But the short duration of this \obs and its large off axis
angle prevented further investigation of the source from this data. Data reduction was performed using CIAO 4.6 using CALDB 4.6.3 and the standard analysis procedure prescribed \footnote{http://cxc.harvard.edu/ciao/}.
The level 1 event file was reprocessed with \textit{chandra$\_$repro}, which incorporates the subpixel repositioning algorithm EDSER as a default for attaining better angular resolution for sources near the centre of the FOV. The effective exposure of the observation after filtering was 38.5 ks. We also checked for possible presence of pileup in the data. Using the tool {\it pileup$\_$map}, the estimated fraction of pileup in the centremost pixel is < $5\%$. Hence the effect is not important for our \obs.

Images were created using the task \textit{dmcopy}. Spatial analysis was performed using the \textit{Sherpa} analysis package 4.4 \footnote{http://cxc.harvard.edu/sherpa4.4/}. The task {\it specextract} was used for extracting source and background spectra and response files from regions of interest.
The selection of regions used for spectral extraction is described in the spectral analysis section. Spectra were fitted using \textit{XSPEC v12.8.1} \footnote{http://heasarc.gsfc.nasa.gov/xanadu/xspec/}.
 
\subsection{Imaging}
\label{sec-im-an}
Figure \ref{images}a shows a full resolution \chandra image (0.5--8 keV) centred on the source near the centre of IKT 16, and zoomed into a region 20\arcs x 25\arcs~ wide. The source appears to be symmetrically elongated in the east-west direction with a bright point source located at its centre. The elongation measures about 5\arcs.
To accurately determine the position of the central source, we created a subpixel image of the same (at $\frac{1}{5}$ of the ACIS pixel resolution) and applied the source detection algorithm {\it celldetect}. The coordinates of the point source are RA(J2000)=$00^h58^m16.85^s$ Dec=$-72\deg18\arcm05.60\arcs$ considering an error of 0.6\arcs~ at 90 $\%$ confidence level in absolute \chandra astrometry. The net count rate from the point source after subtracting the nebular component is 0.011 c/s, and from the entire nebula after subtracting the point source contribution is 0.003 c/s. The details of background subtraction are discussed in the spectral analysis section \ref{sec-spec}. The point source is about three times
brighter than the extended emission. 
After removing the contribution from the bright point source at the centre, and smoothing the image with a Gaussian kernel of width 5\arcsec, there is evidence of a diffuse emission extending further out up to $\sim$ 30$\arcs$ (see Fig.~\ref{images}d). This diffuse component is discussed in Sect.\ref{diff-neb}. The same image shows hints of the presence of the SNR, especially the excess in the north coincident with that in the $\xmmn$ image of IKT 16 \citep{owen2011}.

\subsection{Radio observations}
\label{sec:radioobs}

Our new ATCA observations, project C2521 (CI: J. van Loon), used the Compact Array Broadband Backend (CABB) with the 6A array configuration at 2.1\,GHz providing improved flux density estimates and resolution. These new images were acquired on 02-01-2012 with $\sim10.48$~hours integration over the 12~hour observing session. The radio galaxy PKS 1934-638 was used as a primary flux calibration source for all observations, with the radio sources PKS 0230-790 and PKS 2353-686 used for phase calibration. A standard calibration process was carried out using the \textsc{miriad} data reduction software package (Sault et al. 2011). In order to improve the fidelity and sensitivity of the final image, a single iteration of self-calibration was performed on the strongest sources in the field. A uniform weighting scheme was subsequently used throughout the imaging process, as it provided a balance between improving the theoretical rms noise while maintaining an adequate beam shape. Given the 2\,GHz of bandwidth provided by CABB, images were formed using \textsc{miriad} multi-frequency synthesis (\textsc{mfclean}; Sault \& Wieringa 1994). The same procedure was used for both {\it U} and {\it Q} Stokes parameter maps. However, there was no reliable detection in the {\it U} or {\it Q} intensity parameters associated with this object, implying a lack of polarisation at lower radio frequencies (2.1\,GHz). The final image produced (after primary beam corrected), which possess a FWHM of $3.38\arcsec \times 3.16\arcsec$ and a PA of $-17.8^{\circ}$, has a 1$\sigma$ rms noise level of 16\,$\mu$Jy. In Fig.~\ref{radio} we show the ATCA 2.1\,GHz surface brightness image of PWN IKT\,16.
   
\section{X-ray Spatial analysis}
\label{sec:xspec}
Taking advantage of the excellent spatial resolution of {\it Chandra}, we performed detailed spatial analysis of IKT 16 for the first time. To provide further evidence of the extended emission, and to probe the source extent we created a radial profile centred on the point source from the data.
 We compared this with the two dimensional model of the point spread function (PSF) to look for an excess indicating the extended emission. Further, we tested for signatures of asymmetry along the two halves of the nebular emission. Finally we performed a morphological fitting of the elongated structure with a simple model of the nebula centred on the point source, and determined its geometrical parameters like the size, ellipticity and the rotation angle.
\subsection{Radial profile}
\begin{figure}
\centering
\includegraphics[scale=0.45]{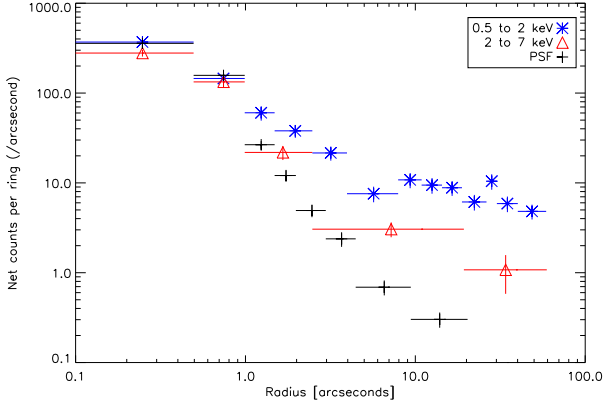}
\caption{Radial counts profile of the source (blue star: 0.5--2 keV; red triangle: 2--7 keV) plotted against the 2-D PSF (0.5--8 keV, in black crosses), clearly showing the extended nature of the source.}
\label{radial}
\end{figure}

Radial profiles were created up to 1$\arcm$ in two energy bands (0.5--2 keV, and 2--7 keV) by extracting net counts in circular annuli centred on the point source using the tool { \it dmextract} and then rebinning to reach a reasonable statistical precision. The background was extracted from a circular region of the same area, away from the source region but inside the SNR.
The PSF of the \obs was simulated
using the \chandra ray tracer chaRT \footnote{http://cxc.harvard.edu/chart/} which simulates the High Resolution Mirror Assembly based on the energy spectrum of the source and the \obs exposure. The output of chaRT was modelled with the software MARX \footnote{http://cxc.harvard.edu/chart/threads/marx/} taking into account the instrumental effects and the EDSER subpixel algorithm to be consistent with the observational data. The best-fit spectrum of the point source (see section \ref{spec-pnt}) was used. Fig.~\ref{radial} shows the radial profile of the \obs along with the simulated PSF. The data at both the energy bands are consistent with a point source up to a radius < 1\arcs~ beyond which it clearly has higher net counts than expected from a point source simulated at the same position and with the same spectral parameters. This corresponds
to the nebular component. 
Although the geometrical model (see section \ref{sec:spa-mod}) indicates that the source FWHM is about 5\arcs, this exercise indicates that the source extends further beyond. This excess is best seen in the energy range of 0.5--2 keV. Its brightness decreases outwards like $r^{-1.5}$. Figure~\ref{images}d indicates that this further extension is probably elongated in the east-west direction like the main X-ray nebula and the radio nebula.
 
\subsection{Investigating signatures of asymmetry in the nebula morphology}
\label{east-west-counts}
Visual inspection of the extended source near the centre of IKT 16 from Fig.~\ref{images}a, gives an impression of an elongated symmetric structure. However, we looked for the possibility of an east-west asymmetry
by dividing the nebular structure into two halves along the axis going through the point source, and compared the total counts from the two regions. The number of counts in the east and west nebula are $50 \pm 9$ and $67 \pm 10$ respectively indicating that the west nebula may be slightly brighter than its other half. It cannot be determined conclusively as the values are consistent within their errors. We investigated this issue further through spectral analysis of the same regions in section \ref{spec-east-west}. 



\begin{table}
\centering
\caption{Parameters of the best-fit model to \pwn in IKT 16.
Model used is a 2-D Gaussian function for the nebula and a constant background.
Theta is the angle between the major axis and the north direction.}
\begin{tabular}{c c c}
\hline

Parameter  & Value & Units\\
 \hline         
FWHM   & 5.2$^{-0.9}_{+1.0}$ & arcsec \\
Ellipticity & 0.6 $\pm 0.1$ & \\
Theta & 82 $\pm 7$ & degrees \\
Amplitude & 2.9 $\pm 1 $ & counts/pixel \\
Background & 0.12 $\pm 0.01$ & counts/pixel \\
\hline    
\end{tabular}
\\
\label{tabsherpa}
\end{table}

\subsection{Spatial modelling}
\label{sec:spa-mod}
Spatial fitting was performed on a larger region of size 35\arcs $\times$ 45\arcs centring the source to detect the existence of a possible diffuse
 emission underlying the nebular one, and also constrain the background better. The PSF described in the 
previous section was loaded as a table model in {\it Sherpa} to model the point source emission. The remaining excess consisted of the background and the emission from the nebula. This was modelled with a constant background and a 2-D unnormalized Gaussian function (gauss2d model in {\it Sherpa}), convolved with the PSF. In the initial iterations it was found that the position of the 2-D Gaussian used to model the nebula was consistent with the point source. In order to constrain the geometrical parameters of the nebula like its FWHM along its major axis, ellipticity and rotation angle better, its position was fixed to the point source henceforth.
The best-fit parameters were determined by the C-statistic \citep{cash1979} and errors were estimated at 90$\%$ confidence level. The residuals do not show any systematic pattern indicating that this analysis is not sensitive to any substructures apart from the elongated structure of the nebula. Fig.~\ref{images}b shows the model with the contours from the data smoothed with a Gaussian of $\sigma$ = 3\arcs overlayed on it. This highlights the fact that the model is a good description of the data. It is worth mentioning that we also tried to fit a 2-D Lorentz model with a varying power-law (beta2d model in {\it Sherpa}) instead of the 2-D Gaussian used to model the nebula. We noticed that by freezing the rotation angle and ellipticity, the power-law index $\alpha$ tends towards the maximum limit of 10 which closely approximates the 2-D Gaussian model. The difference in C-statistic between beta2d and gauss2d is 15 in the Lorentzian limit ($\alpha$=1 in beta2d), and 2 in the Gaussian limit ($\alpha$=10 in beta2d). We therefore concluded that the data does not favor broad wings and a 2-D Gaussian is preferred for the nebular emission.

The best fit parameters are listed in Table \ref{tabsherpa}. The estimated FWHM of 5.2\arcs~is in agreement with that measured from the image of the source. It is inclined at an angle of 82\deg~ between the long axis and the north direction and is aligned with the radio nebula (see Fig.~\ref{radio}). 
An ellipticity of 0.6 indicates a major to minor axis ratio of 0.8. Another important result is that there is no signature of displacement between the point source and the nebula.
The best-fit position of the centre of the nebula obtained corresponds to R.A.(J2000)=00:58:16.824 and DEC.(J2000)=-72.18:05.32 with an error of 0.1\arcs and 0.2\arcs on the x and y positions respectively. This is consistent with the position of the point source obtained with {\it celldetect }.


\section{X-ray Spectral analysis}
\label{sec-spec}
We have performed detailed spectral analysis of the point source and the nebular component of IKT 16. In addition we have also investigated the diffuse component extending to larger scales.
The main deciding factors for the spectral analysis are the regions used for the spectral extraction and the background modelling, taking into account the contribution
of the point source in the nebular spectrum and vice versa. This is described in the subsequent subsections. The analysis was performed in the energy range of 0.5--7 keV. C-statistic was used for spectral fitting and errors were estimated at 90$\%$ confidence interval. To account for the photoelectric absorption by the interstellar gas, two absorption components were used as in \cite{owen2011}. The first, {\it phabs} component was fixed at the
Galactic value of $6\times10^{20}\ergsec$ (\citealt{dickey90}), and a free absorption ({\it vphabs}) component was used to account for absorption inside the SMC. This second component has metal abundances fixed at 0.2 solar, as is typical in the SMC (\citealt{russell92}). 
\subsection{Point source spectrum} 
\label{spec-pnt}
For the point source, a circular region of radius 1.5 pixels was extracted centred on the best-fit coordinates of the source. The extraction region is shown in Fig.
\ref{images}c. A circular annulus outside the nebular extraction region was used for the background spectrum. Apart from this the
astrophysical background due to the nebular contribution was also considered. For this we calculated the fraction of nebular emission contributing in the point source extraction region (from the best-fit morphological model of the source), and accounted for it as an additional model component in the spectral fitting, with parameters fixed to the best fit values obtained in section \ref{spec-pwn}.

\begin{figure*}
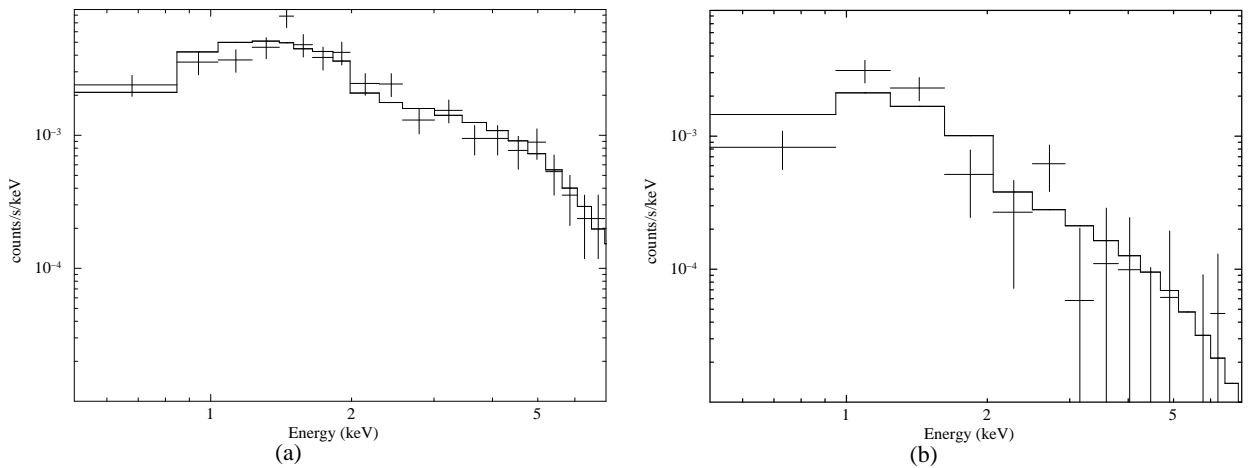

\centering
\subfigure[]{\includegraphics[height=0.45\textwidth,angle=-90]{pt_src_without_resi.ps}}
\subfigure[]{\includegraphics[height=0.45\textwidth,angle=-90]{total_neb_new_resi.ps}}
 \caption{The figures show the ACIS-S spectra for the putative pulsar (a,{\it left}) and the entire PWN (b,{\it right}).
The solid lines correspond to the respective best-fit spectral models. The plots have been rebinned for visual clarity.}
\label{spec-1}
\end{figure*}

We tried to fit the spectrum with several models including an absorbed power-law, an absorbed blackbody and a combination of both models. The absorbed blackbody model leads to an unacceptable spectral fit with a difference of 54 in C-statistic value between the two models with respect to the power-law model. The absorbed power-law model provides a good fit as illustrated on Fig.~\ref{spec-1}a. The best fit parameters along with the flux are tabulated in Table \ref{table-specfit}. The addition of a blackbody component was not required and did not improve the fit significantly. Also, the additional absorption column density inside the SMC could not be constrained well and was prone to large error bars as can be seen from Table \ref{table-specfit}. Its value was however
consistent with that obtained from the previous analysis of the \xmmn \obss by \cite{owen2011} which had smaller errors associated with it probably due to the larger number of counts. 
\subsection{Nebula spectrum}
\label{spec-pwn}
For the spectral extraction of the extended source, a box region was used with the region corresponding to the point source excised (see Fig.~\ref{images}c).
Its size was optimized from the morphological modelling of the nebula, and its centre was made coincident with the point source. The background was taken from the same annular region that was used as a background for the point source spectral extraction. For the additional astrophysical background contributed by the point source leakage in the nebular region,
we extracted a spectrum from the events from the PSF simulated by chaRT with spectral parameters similar to those of the point source, in the same region that was used for the nebular spectral
extraction, and added it to the internal background spectrum as an additional background component. 

We used an absorbed power-law as the spectral model for all the nebular fits described henceforth. In contrast to the point source spectrum, the spectrum of the nebula is very soft. Keeping the absorption inside the SMC free in these fits leads to strong correlation between the absorption column density and the power-law index $\Gamma$. To avoid this it is better to constrain this parameter from the point source spectrum, and freeze it to this value for the nebular spectral fits. However, the better capability of the \xmm \obss of IKT 16 to constrain the local absorption density, as discussed earlier, led us to fix this value to that obtained from the previous analysis of IKT 16 with \xmmn\citep{owen2011} in all the nebular fits. The spectrum along with its best-fit model and residuals is shown in Fig.~\ref{spec-1}b, and the best fit parameters are tabulated in Table \ref{table-specfit}.
\subsection{Outer and Inner nebula spectra}
\label{spec-pwn-inner}
In order to look for changes in the spectral parameters, particularly a spectral steepening with radius in the nebula, we divided the total nebular region into two parts, by choosing an inner annular region centred on the point source, and an outside region excluding it. The extraction regions are shown in Fig.~\ref{images}c. The Fig.~\ref{spec-2}a and \ref{spec-2}b show the inner and outer spectrum respectively with their best-fit models and residuals, and Table \ref{table-specfit} their best-fit parameters. Although the outer nebula is hinted to be softer then its inner counterpart by the absence of counts above 5 keV, we detect no steepening of the spectral index $\Gamma$ which would be indicative of synchrotron cooling. The $\Gamma$ values are consistent within errors at 90 $\%$ confidence. 
\subsection{East-west nebula spectra}
\label{spec-east-west}
\begin{figure*}[htp]
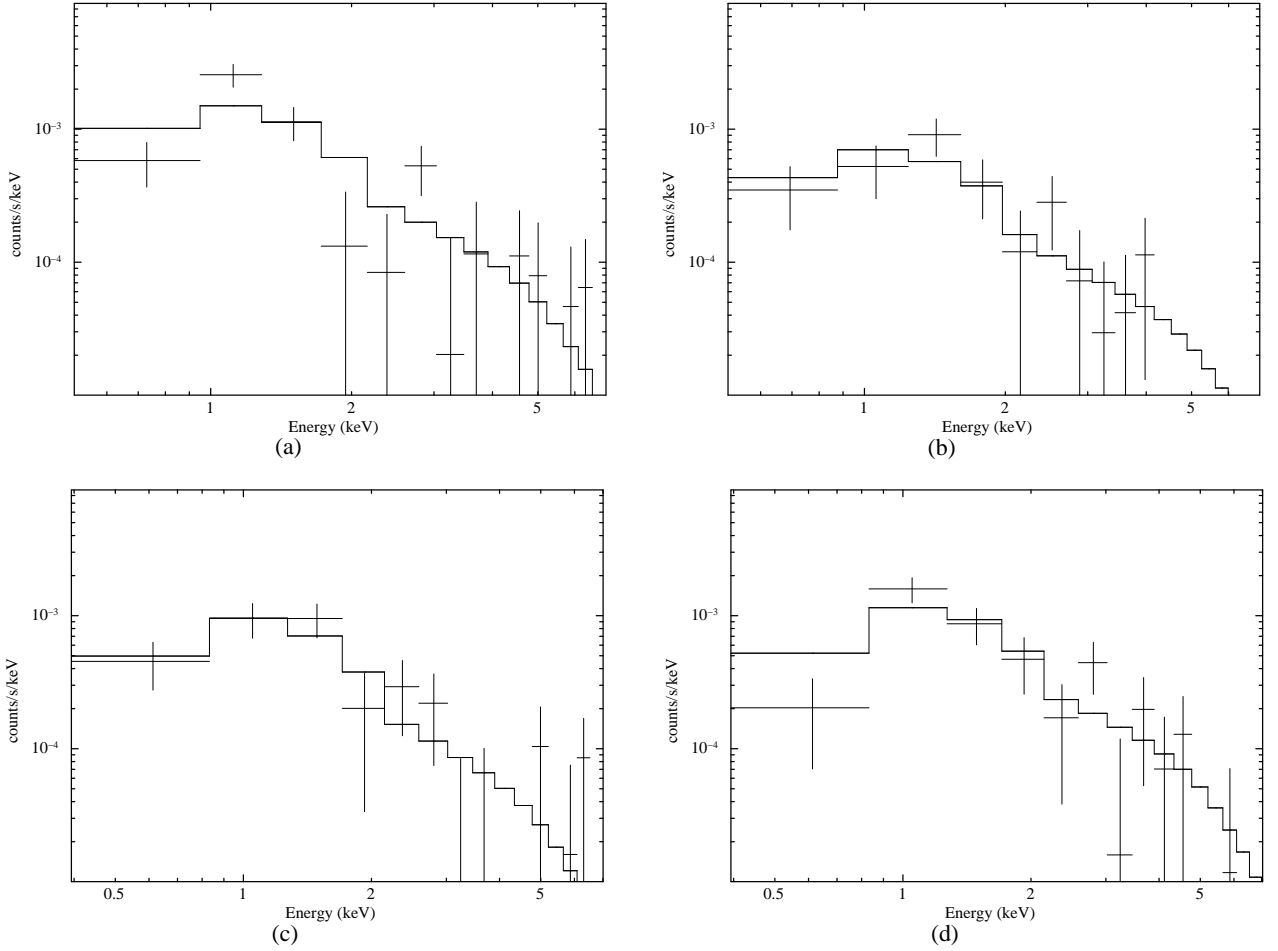

\centering
\subfigure[]{\includegraphics[height=0.45\textwidth,angle=-90]{inner_new_resi.ps}}\quad
\subfigure[]{\includegraphics[height=0.45\textwidth,angle=-90]{outer_new_resi.ps}} \\
\subfigure[]{\includegraphics[height=0.45\textwidth,angle=-90]{nebula_east_spec_new_resi.ps}} \quad
\subfigure[]{\includegraphics[height=0.45\textwidth,angle=-90]{nebula_west_spec_new_resi.ps}} \\
\caption{Same as in Fig.~\ref{spec-1} but for the inner nebula {\it a}, outer nebula {\it b}, east nebula {\it c} and west nebula {\it d}. Please note that in the outer nebula, i.e. Fig.~{\it b}, there are no events above 5 keV.}
\label{spec-2}
\end{figure*}

In Sect.~\ref{east-west-counts} where we report the extracted total number of counts in the east and west nebular region, we find evidence that the west of the nebula may be slightly brighter although the values are consistent within errors. In continuation, we also extract spectra from the same regions following the same procedure as in Secs.   \ref{spec-pwn} and \ref{spec-pwn-inner}. Figures~\ref{spec-2}c and Fig.~\ref{spec-2}d show the east and west spectrum respectively with their best-fit models and residuals, and Table \ref{table-specfit} their best-fit parameters. As indicated by the earlier exercise in Sect. \ref{east-west-counts}, the spectrum from the west of the nebula shows slightly higher value of flux but it is comparable to that measured from the east nebula within errors at 90 $\%$ confidence level. We do not detect any significant change in the spectral index $\Gamma$. 
\subsection{Diffuse emission}
\label{diff-neb}
From Figs.~\ref{images}d and \ref{radial}, we see evidence that the source extends further beyond the region adopted to study the nebula spectrum. To investigate further, we extracted the spectrum of the diffuse component by choosing an annular region excluding the nebular extraction region (box region in Fig.~\ref{images}c) and extending up to the size of the \xmmn point source extraction region of 20\arcs \citep{owen2011}.
Using the same spectral model as that for the nebular spectrum, we find that the diffuse component has a flux and spectral index (tabulated in Table \ref{table-specfit}) comparable to that of the main nebular component. We also tried to fit the spectrum with a Sedov model, since that emission might be associated with the SNR. We fixed the parameters to those obtained by \cite{owen2011} (the data quality does not allow fitting anything else than the normalisation). The resulting C-statistic is higher by 9 than that of the power-law model. This does not favor the thermal model, but does not allow ruling it out either. However, no other part of the SNR is as bright, and this diffuse emission is centred on the main X-ray nebula. So we favor the interpretation as a further PWN component.

In addition, we also investigated whether this emission could be a halo due to the scattering from the foreground dust in the interstellar medium. \cite{predel1995} using \emph{ROSAT} observations, studied X-ray scattering halos around 25 point sources including the three bright sources in the LMC, LMC X-1, LMC X-2 and LMC X-3. The local absorbing density around IKT 16 is lower that in LMC X-1, but higher than in LMC X-2 and LMC X-3, and thus lies in the range covered in this study. The authors found that the relative intensity of the scattering halos of the LMC sources w.r.t the point source are $\sim$ 1\%. This is much weaker than what we find for the diffuse component which is $\sim$ 25\% (< 2 keV) of the putative pulsar emission. Moreover, \emph{ROSAT} observed a flat profile for the halos extending to > 100\arcs. Within our extraction region of the diffuse emission (20\arcs), the fraction is expected to be even lower than 1\%. Hence the dust scattering origin for this diffuse emission is very unlikely.

Finally, we compared our results with the spectral model of the point source seen with \xmmn \citep[Table 3 in][counting the spillover of the point source into the SNR extraction region, and the SNR emission in the point source region]{owen2011}. The total flux measured in the energy range of 0.5--8 keV from the \chandra observation is $21.3 \times 10^{-14}$
$\ergcms$ 
(putative pulsar + nebula + diffuse emission; see Table \ref{table-specfit} for the obtained values) comparable to $24.9 \times 10^{-14}$ $\ergcms$ measured from the \xmmn spectral model. The unabsorbed luminosity in the same energy range (0.5--8 keV) measured  from \chandra~ is \textbf{1.0} $\times10^{35}$ $\ergsec$. In comparison, the \xmm spectral model corresponds to an unabsorbed luminosity of 1.1 and 1.3 $\times10^{35}$ $\ergsec$ in the 0.5--8 keV and 0.5--10 keV energy ranges and is in agreement with the \chandra results. It should be noted that, although the total unabsorbed luminosity (0.5--10 keV, SNR + point source) in \cite{owen2011} is consistent with the spectral model, there seems to be an error in quoting the unabsorbed luminosity of the point source as 1.6 $\times10^{35}$ $\ergsec$ in that paper.

\begin{table*}
\caption{Parameters of the best-fit spectral model to putative pulsar and the nebula. Errors are quoted at 90 $\%$ confidence.
The \chandra energy range is 0.5--8 keV.}
\centering
\begin{tabular}{lcccccc}
\hline

Region  & SMC absorption &  Power-law photon Index  & \fx$^{a}$ &  \fx$^{a}$ &  \lx$^{b}$\\
   & $10^{21}$cm$^{-2}$  &  $\Gamma$  & 0.5--8 keV &  2--10 keV &  0.5--8 keV \\
\hline
Putative pulsar  & 5.3$^{+3.9}_{-3.5}$ & 1.11$\pm0.23$ & 16$\pm 1$  & 17$\pm2$ & 7.20\\
Nebula & 3.4 (f)$^{c}$  & 2.21$^{+0.40}_{-0.37}$ & 2.2$\pm0.2$  & 1.3$^{+0.4}_{-0.5}$ & 1.19 \\
Inner nebula & 3.4 (f)$^{c}$ & 2.22$^{+0.48}_{-0.45}$ & 1.6$\pm0.3$  & 1.0$^{+0.5}_{-0.7}$ & 0.78\\
Outer nebula & 3.4 (f)$^{c}$ & 2.18$^{+0.60}_{-0.54}$ & 0.75$^{+0.38}_{-0.41}$ & 0.47$\pm0.63$ & 0.41\\
West nebula & 3.4 (f)$^{c}$  & 1.97$^{+0.52}_{-0.40}$  & 1.4$\pm 0.3$ & 0.9$^{+0.5}_{-0.6}$ & 0.67 \\
East nebula & 3.4 (f)$^{c}$  & 2.23$^{+0.67}_{-0.47}$ & 0.93$^{+0.4}_{-0.36} $  & 0.55$\pm0.67$ & 0.45 \\
Diffuse emission & 3.4 (f)$^{c}$ & 2.20$^{+0.38}_{-0.30}$ & 3.1$\pm0.2$ & 1.9$^{+0.3}_{-0.2}$ & \textbf{1.70} \\
\hline 
\label{tabspec}   
\end{tabular}
\\
$^{a}$ - Observed flux in units of $10^{-14}$ $\ergcms$ \\
$^{b}$ - absorption corrected luminosity in units of $10^{34}$ $\ergsec$ assuming a distance of \textbf{61} kpc. \\
(f) - Parameter fixed for consistency between fit regions. \\ 
$^{c}$ - SMC absorption for all nebular spectra was frozen to the column density obtained from the \xmmn \obs. \\
\label{table-specfit}
\end{table*}
\section{Radio}
The PWN in the radio-continuum regime extends far beyond X-ray detection as can be seen in Fig.~\ref{radio}. The radio extent of the source as measured visually, is $70\pm10$\arcs.
We note that the X-ray point source correlates to one of the peaks in our RC image. However, no corresponding point source can be drawn anywhere near the X-ray point source.

We point to the new high resolution data (Fig.~\ref{radio}) where two lobe like features, somewhat symmetric, each $20\pm5$\arcs~in extent, outside of the X-ray nebula are seen. These radio-continuum features oppose our initial scenario that the radio nebula has been pushed aside by the reverse shock \citep{owen2011}. It seems it is more likely that we are still seeing the nebular expansion in the cold ejecta. 

Across the 2.1\,GHz band, the radio spectral index is flat as reported in \cite{owen2011}. Somewhat surprising, we did not detect any polarisation in either {\it Q}, {\it U} or {\it V} stokes parameters even that our detection level is better than 1\%. The small scale structure may play an important role in smoothing out any weak polarisation. At the same time, depolarisation may reveal regions of underlying turbulence and/or compression and heating of thermal material at various shocks within the remnant system \citep{anderson1995}. This depolarisation may also indicate somewhat older age of the remnant. The noticeably younger PWNe in the LMC such as N157B \citep{lazendic}, SNR J0453-6829 \citep{hab2012}, and 0540-69.3 \citep{bran2014} all have strong polarisation. 
\section{Discussion}
\label{sec:disc}
Confirmation of the PWN nature of the source makes IKT 16 the first composite SNR to be discovered in the SMC.
The nebular structure is consistent with the presence of an elongated PWN around a putative pulsar, which extends further out in the form of a fainter diffuse emission.
However, we cannot rule out the presence of an inner compact X-ray nebula of the order of 0.1 pc which cannot be resolved at the distance of the SMC. 
Therefore, the overall impression is that of the bright central point source corresponding to a putative pulsar or a putative pulsar + unresolved inner PWN component, and an  extended emission component representing the PWN.

 Although the present \chandra \obs in full-frame mode is not capable of detecting pulsations from the central putative pulsar, we can nevertheless infer its properties to a certain extent by assuming some basic characteristics of the neutron star as is described below. In the later subsections, we also discuss the evolutionary stage of the \pwn as expected from its morphology.
\subsection{Derived properties of the pulsar}
\label{sec:discsnr}

\begin{figure}
\centering 
\includegraphics[height=0.45\textwidth,angle=-90]{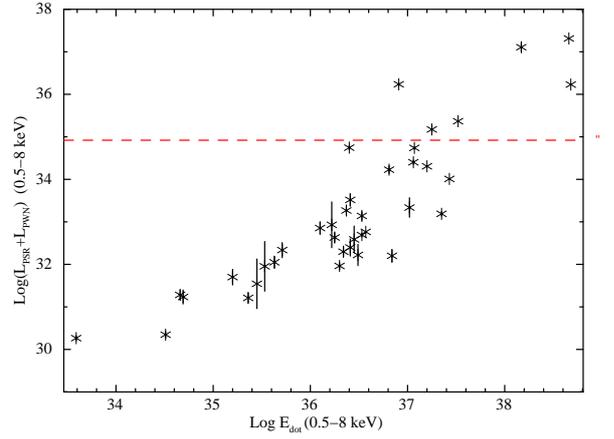} 
\caption{ Dependence of the total non-thermal luminosity (nebula+pulsar) on the pulsar spin-down power for the PWNe observed with \chandra. Adapted from \cite{pwnchandra}. The red dashed line indicates the luminosity of the putative pulsar + PWN in IKT 16.}
\label{corr}
\end{figure}

The nebular emission in X-rays reflects the youngest generation of the emitting particles. Therefore, a strong correlation is expected 
between the non-thermal X-ray luminosity of the PWN and its pulsar, with the spin down properties of the
pulsar itself. The properties of the putative pulsar in IKT 16 were estimated adopting the work of \cite{pwnchandra}. In this paper the correlation between the luminosity of the PWN, the non-thermal luminosity of the pulsar, and its various properties are derived from a large sample of \chandra \obss of PWNe. The unabsorbed luminosities of the PWN and the pulsar in IKT 16 are reported in Table \ref{tabspec}. Using these values and the tabulated properties of the PWNe from \cite{pwnchandra}, we estimated the spin-down power of the putative pulsar by comparing the total non-thermal luminosity of the sources (PWN+pulsar component) with their spin-down power. This ensured our estimate was free from the assumption that the central source is exclusively the pulsar and does not contain a compact nebular component. It is noteworthy to mention that in order to perform this estimation, we have not accounted for the flux from the diffuse emission (described in section \ref{diff-neb}) which is not as well measured and whose origin is less unambiguous. At the same time addition of this component would not alter our results in the limits of the correlation measured in \cite{pwnchandra}. Fig.~\ref{corr} plots the spin-down power against the total luminosity for all the objects from \cite{pwnchandra}, with a line drawn to indicate the total luminosity of the central source in IKT 16. It corresponds to an expected spin-down power $\dot{E} \sim 10^{37} \ergsec$.  Further, assuming its age to be the same as that measured from the SNR \citep{owen2011}, $\ie$ 14.7 kyr, we overlayed the central source on the $P - \dot{P}$ diagram from the ATNF catalogue \citep{atnf}. The expected spin period is < 100 ms with $\dot{P} \sim 10^{-13}$ s s$^{-1}$. This points towards a young and energetic pulsar powering the PWN in IKT 16, in fact the youngest in the SMC. Future on-axis \xmm \obss or \chandra \obss using operating modes with better timing resolution can be used to search for pulsations from this source.
\subsection{Energy spectrum}
\label{discspec}
The spectra of both the central and the extended source can be satisfactorily fitted with a power-law model, expected from a synchrotron emission dominated spectrum. The hard power-law spectral index 1.1 for the point source resembles that of a typical young rotation powered pulsar where the non-thermal radiation is generated by the particles accelerated in the pulsar magnetosphere \citep[][and references therein]{michel1992}.
No additional thermal component (which would denote the emission from the surface or the polar caps of the neutron star) was required to fit the pulsar spectrum. The point source is  also three times brighter than the surrounding nebula. If this comprises exclusively the pulsar emission, it is rather unusual since the average value of the nebular to pulsar emission is $\langle\frac{\eta_{pwn}}{\eta_{psr}}\rangle \sim 4 $ for a large sample of PWNe observed with \chandra \citep{pwnchandra}. The $\Gamma$ of the nebular emission is compatible with that usually observed for PWNe \citep[1.5 to 2.1:][]{pwnchandra,li2008}, and is also consistent with particle wind models of Fermi shock acceleration \citep{2001MNRAS.328..393A}.
\subsection{Morphology}
\label{sec:discsnrmorp}
The morphology of the PWN is elongated along the east-west direction as reported in section \ref{spec-east-west}. 
The obtained size of  5.2\arcs~of the X-ray nebula corresponds to a source extent of 1.5 pc at the distance of the SMC. The radio extent of the source is much larger corresponding to 21 pc with the brightest emission from the radio lobes, each $\sim$ 6 pc in extent assuming the same distance.
This scenario is consistent with a longer cooling time for radio emission compared to that of X-ray emission.
In addition, the radio nebula extends even further than the estimated SNR centre, consistent with the fact that the earliest electrons were produced when the pulsar was much closer to the SNR centre. Further, the long axis of the radio and X-ray nebula are aligned with each other. This axis of alignment points to the centre of the SNR, so is presumably aligned with the pulsar's direction of motion.
\subsection{Interaction of the SNR with the PWN  and its stage of evolution}
\label{pwn-interact}
The SNR in IKT 16 is in the adiabatic Sedov phase of evolution with an age of $\sim$ 14.7 kyr \citep{owen2011}. The shell of the SNR has important implications on the PWN, and the physics of these systems is extremely complicated due to the rapid evolution of both the SNR and the central pulsar \citep{gaensler2006,gelfand2007,gelfand2009}. The evolution of a PWN inside an SNR follows three important evolutionary phases: an initial free-expansion in the supernova ejecta, the collision between the PWN and SNR reverse shock, which crushes the PWN subjecting it to various instabilities, and eventually subsonic re-expansion of the PWN in the shock heated ejecta \citep[][and references therein]{2001ApJ...563..806B,2003A&A...397..913V,2003A&A...404..939V,2004A&A...420..937V,gelfand2007}. After the reverse shock interaction has died off, $\ie$ in the PWN re-expansion stage, the neutron star is often displaced from its PWN, leaving behind the "relic" PWN usually observed in radio and forms a "new" PWN comprising of freshly injected particles, and therefore observable in the X-rays. A later additional phase of "bow-shock" nebula can sometimes be identified, as the local sound velocity decreases with the pulsar's progression through the SNR. In this phase the pulsar's motion may become supersonic when it approaches the shell of the SNR, and it may acquire a bow-shock morphology \citep{2003A&A...397..913V,2003A&A...404..939V,2004A&A...420..937V}.

Adopting the total mechanical energy released in the explosion ($E_{51}$ in units of $10^{51}$ erg) and the ambient medium density ($n_{0}$) from the adiabatic Sedov modeling in \cite{owen2011}, and assuming the mass of the ejecta, $M_{ej} = 10 M_{\odot}$, we can calculate the reverse shock trajectory \citep{mckee1995} at the present age of the SNR (14.7 kyr). At the distance of the SMC, we obtain that the reverse shock radius $R_{s}$ is 11 pc away from the centre of the SNR. This would imply that although it might not have encountered the smaller X-ray nebula yet \citep[measured to be at 8 pc from the centre of the SNR][]{owen2011}, it would have crossed the larger radio extent of the nebula and interacted with the lobes. This picture is however not supported by the new high resolution radio data. The presence of two bright radio lobes more or less symmetric about the X-ray counterpart is consistent with an expansion of the PWN into cold ejecta in homologous expansion. 
We however point out that there may be uncertainties in measurement of our Sedov model parameters for the SNR IKT 16 \citep{owen2011} which may limit the estimation of $R_{s}$.
\section{Conclusions}
\label{sec:conc}
Thanks to the high resolution \chandra ACIS \obs we have been able to resolve the hard X-ray emission near the centre of the SNR IKT 16 into a PWN centred on a
putative pulsar. We have imaged the PWN, constrained its geometrical parameters, and have measured all the  spectral components separately. With new high resolution radio data we have determined the radio morphology and its extent precisely. The main results can be summarized as follows:
\begin{itemize}
\item The putative PWN is elongated but centred on the point source. Comparison with the radio counterpart indicates that the reverse shock may not have yet met with the PWN surface.
\item The point source at the centre is about three times brighter than the elongated feature.
\item Morphological modelling of the X-ray nebula with a PSF and a 2D Gaussian resulted in an FWHM of 5.2\arcs~for the PWN with its axis aligned with the larger radio nebula.
\item The point source at the centre has a much harder spectrum than the extended emission from the nebula. This points to the presence of a 
pulsar dominated by non-thermal emission. The expected $\dot{E}$ is $\sim 10^{37} \ergsec$ and spin period < 100 ms, with  
$\dot{P} \sim 10^{-13}$ s s$^{-1}$. This points towards a young and energetic pulsar powering the PWN in IKT 16.
However, the presence of a compact X-ray PWN unresolved by Chandra at the distance of the SMC cannot be ruled out.
\end{itemize} 


\begin{acknowledgements}
This research has made use of data obtained from the \chandra X-ray observatory. We used the KARMA and MIRIAD software packages developed by the
ATNF. The Australia Telescope Compact Array is part of the Australia
Telescope which is funded by the Commonwealth of Australia
for operation as a National Facility managed by CSIRO. CM acknowledges Fabio Acero for very useful discussions and comments on the paper. The authors would like to acknowledge the referee Satoru Katsuda for very useful comments which significantly improved the paper.
\end{acknowledgements}

\bibliography{ikt16}

\begin{thebibliography}{26}
\expandafter\ifx\csname natexlab\endcsname\relax\def\natexlab#1{#1}\fi

\bibitem[{{Achterberg} {et~al.}(2001){Achterberg}, {Gallant}, {Kirk}, \&
  {Guthmann}}]{2001MNRAS.328..393A}
{Achterberg}, A., {Gallant}, Y.~A., {Kirk}, J.~G., \& {Guthmann}, A.~W. 2001,
  \mnras, 328, 393

\bibitem[{{Anderson} {et~al.}(1995){Anderson}, {Keohane}, \&
  {Rudnick}}]{anderson1995}
{Anderson}, M.~C., {Keohane}, J.~W., \& {Rudnick}, L. 1995, \apj, 441, 300

\bibitem[{{Blondin} {et~al.}(2001){Blondin}, {Chevalier}, \&
  {Frierson}}]{2001ApJ...563..806B}
{Blondin}, J.~M., {Chevalier}, R.~A., \& {Frierson}, D.~M. 2001, \apj, 563, 806

\bibitem[{{Brantseg} {et~al.}(2014){Brantseg}, {McEntaffer}, {Bozzetto},
  {Filipovic}, \& {Grieves}}]{bran2014}
{Brantseg}, T., {McEntaffer}, R.~L., {Bozzetto}, L.~M., {Filipovic}, M., \&
  {Grieves}, N. 2014, \apj, 780, 50

\bibitem[{{Cash}(1979)}]{cash1979}
{Cash}, W. 1979, \apj, 228, 939

\bibitem[{{Crawford} {et~al.}(2014){Crawford}, {Filipovi{\'c}}, {McEntaffer},
  {Brantseg}, {Heitritter}, {Roper}, {Haberl}, \& {Uro{\v
  s}evi{\'c}}}]{crawford2014}
{Crawford}, E.~J., {Filipovi{\'c}}, M.~D., {McEntaffer}, R.~L., {et~al.} 2014,
  \aj, 148, 99

\bibitem[{{Curtis}(1992)}]{michel1992}
{Curtis}, M. 1992, Ciel et Terre, 108, 32

\bibitem[{{Dickey} \& {Lockman}(1990)}]{dickey90}
{Dickey}, J.~M. \& {Lockman}, F.~J. 1990, \araa, 28, 215

\bibitem[{{Evans} {et~al.}(2010){Evans}, {Primini}, {Glotfelty}, {Anderson},
  {Bonaventura}, {Chen}, {Davis}, {Doe}, {Evans}, {Fabbiano}, {Galle}, {Gibbs},
  {Grier}, {Hain}, {Hall}, {Harbo}, {(Helen He}, {Houck}, {Karovska},
  {Kashyap}, {Lauer}, {McCollough}, {McDowell}, {Miller}, {Mitschang},
  {Morgan}, {Mossman}, {Nichols}, {Nowak}, {Plummer}, {Refsdal}, {Rots},
  {Siemiginowska}, {Sundheim}, {Tibbetts}, {Van Stone}, {Winkelman}, \&
  {Zografou}}]{evans2010}
{Evans}, I.~N., {Primini}, F.~A., {Glotfelty}, K.~J., {et~al.} 2010, \apjs,
  189, 37

\bibitem[{{Gaensler} \& {Slane}(2006)}]{gaensler2006}
{Gaensler}, B.~M. \& {Slane}, P.~O. 2006, \araa, 44, 17

\bibitem[{{Gelfand} {et~al.}(2007){Gelfand}, {Gaensler}, {Slane}, {Patnaude},
  {Hughes}, \& {Camilo}}]{gelfand2007}
{Gelfand}, J.~D., {Gaensler}, B.~M., {Slane}, P.~O., {et~al.} 2007, \apj, 663,
  468

\bibitem[{{Gelfand} {et~al.}(2009){Gelfand}, {Slane}, \& {Zhang}}]{gelfand2009}
{Gelfand}, J.~D., {Slane}, P.~O., \& {Zhang}, W. 2009, \apj, 703, 2051

\bibitem[{{Haberl} {et~al.}(2012{\natexlab{a}}){Haberl}, {Filipovi{\'c}},
  {Bozzetto}, {Crawford}, {Points}, {Pietsch}, {De Horta}, {Tothill}, {Payne},
  \& {Sasaki}}]{hab2012}
{Haberl}, F., {Filipovi{\'c}}, M.~D., {Bozzetto}, L.~M., {et~al.}
  2012{\natexlab{a}}, \aap, 543, A154

\bibitem[{{Haberl} {et~al.}(2012{\natexlab{b}}){Haberl}, {Sturm}, {Ballet},
  {Bomans}, {Buckley}, {Coe}, {Corbet}, {Ehle}, {Filipovic}, {Gilfanov},
  {Hatzidimitriou}, {La Palombara}, {Mereghetti}, {Pietsch}, {Snowden}, \&
  {Tiengo}}]{haberl2012}
{Haberl}, F., {Sturm}, R., {Ballet}, J., {et~al.} 2012{\natexlab{b}}, \aap,
  545, A128

\bibitem[{{Kargaltsev} \& {Pavlov}(2008)}]{pwnchandra}
{Kargaltsev}, O. \& {Pavlov}, G.~G. 2008, in American Institute of Physics
  Conference Series, Vol. 983, 40 Years of Pulsars: Millisecond Pulsars,
  Magnetars and More, ed. C.~{Bassa}, Z.~{Wang}, A.~{Cumming}, \& V.~M.
  {Kaspi}, 171--185

\bibitem[{{Lazendic} {et~al.}(2000){Lazendic}, {Dickel}, {Haynes}, {Jones}, \&
  {White}}]{lazendic}
{Lazendic}, J.~S., {Dickel}, J.~R., {Haynes}, R.~F., {Jones}, P.~A., \&
  {White}, G.~L. 2000, \apj, 540, 808

\bibitem[{{Li} {et~al.}(2008){Li}, {Lu}, \& {Li}}]{li2008}
{Li}, X.-H., {Lu}, F.-J., \& {Li}, Z. 2008, \apj, 682, 1166

\bibitem[{{Manchester} {et~al.}(2005){Manchester}, {Hobbs}, {Teoh}, \&
  {Hobbs}}]{atnf}
{Manchester}, R.~N., {Hobbs}, G.~B., {Teoh}, A., \& {Hobbs}, M. 2005, \aj, 129,
  1993

\bibitem[{{McKee} \& {Truelove}(1995)}]{mckee1995}
{McKee}, C.~F. \& {Truelove}, J.~K. 1995, \physrep, 256, 157

\bibitem[{{Owen} {et~al.}(2011){Owen}, {Filipovi{\'c}}, {Ballet}, {Haberl},
  {Crawford}, {Payne}, {Sturm}, {Pietsch}, {Mereghetti}, {Ehle}, {Tiengo},
  {Coe}, {Hatzidimitriou}, \& {Buckley}}]{owen2011}
{Owen}, R.~A., {Filipovi{\'c}}, M.~D., {Ballet}, J., {et~al.} 2011, \aap, 530,
  A132

\bibitem[{{Predehl} \& {Schmitt}(1995)}]{predel1995}
{Predehl}, P. \& {Schmitt}, J.~H.~M.~M. 1995, \aap, 293, 889

\bibitem[{{Russell} \& {Dopita}(1992)}]{russell92}
{Russell}, S.~C. \& {Dopita}, M.~A. 1992, \apj, 384, 508

\bibitem[{{van der Heyden} {et~al.}(2004){van der Heyden}, {Bleeker}, \&
  {Kaastra}}]{van2004}
{van der Heyden}, K.~J., {Bleeker}, J.~A.~M., \& {Kaastra}, J.~S. 2004, \aap,
  421, 1031

\bibitem[{{van der Swaluw}(2003)}]{2003A&A...404..939V}
{van der Swaluw}, E. 2003, \aap, 404, 939

\bibitem[{{van der Swaluw} {et~al.}(2003){van der Swaluw}, {Achterberg},
  {Gallant}, {Downes}, \& {Keppens}}]{2003A&A...397..913V}
{van der Swaluw}, E., {Achterberg}, A., {Gallant}, Y.~A., {Downes}, T.~P., \&
  {Keppens}, R. 2003, \aap, 397, 913

\bibitem[{{Van der Swaluw} {et~al.}(2004){Van der Swaluw}, {Downes}, \&
  {Keegan}}]{2004A&A...420..937V}
{Van der Swaluw}, E., {Downes}, T.~P., \& {Keegan}, R. 2004, \aap, 420, 937

\end{thebibliography}

\end{document}